\documentclass[12pt,letterpaper,english]{article}
\usepackage[T1]{fontenc}
\usepackage[latin1]{inputenc}
\usepackage{babel}
\usepackage{graphics}

\makeatletter

\providecommand{\LyX}{L\kern-.1667em\lower.25em\hbox{Y}\kern-.125emX\@}
\newcommand{\noun}[1]{\textsc{#1}}

 \newcommand{\lyxaddress}[1]{
   \par {\raggedright #1 
   \vspace{1.4em}
   \noindent\par}
 }

\usepackage[T1]{fontenc}
\usepackage{setspace}


\makeatother
\begin{document}

\title{On the origin of plankton patchiness }

\author{J. M. G. Vilar\( ^{*} \), R. V. Sol\'{e}\( ^{\dagger } \), and
J. M. Rub\'{\i}\( ^{\ddagger } \)}

\maketitle

\lyxaddress{\( ^{*} \)Center for Studies in Physics and Biology, The Rockefeller
University, 1230 York Avenue, New York, New York 10021, USA }

\lyxaddress{\( ^{\dagger } \)Department of Physics, FEN, Universitat Polit\'{e}cnica
de Catalunya, Campus Nord, M\`{o}dul B4, E-08034 Barcelona, Spain}

\lyxaddress{\( ^{\ddagger } \)Departament de F\'{\i}sica Fonamental, Facultat
de F\'{\i}sica, Universitat de Barcelona, Diagonal 647, E-08028 Barcelona,
Spain}

\vfill



\newpage

~\\
\textbf{Abstract}\\

Plankton is the productive base of aquatic ecosystems and plays a
major role in the global control of atmospheric carbon dioxide. Nevertheless,
after intensive study, the factors that drive its spatial distribution
are still far from being clear. The models proposed so far show very
limited agreement with actual data as many of their results are not
consistent with field observations. Here we show that fluctuations
and turbulent diffusion in standard prey-predator models are able
to accurately and consistently explain plankton field observations
at mesoscales (1-100 km). This includes not only the spatial pattern
but also its temporal evolution. We explicitly elucidate the interplay
between physical and biological factors, suggesting that the form
in which small scale biotic fluctuations are transferred to larger
scales may constitute one of the key elements in determining the spatial
distribution of plankton in the sea. 

\newpage

~\\
Understanding how complex ecosystems work often relies on simplified
models that disregard many details of the actual system while retaining
the essential information~\cite{levi1,murray,JB1}. In the case of
marine ecosystems, not only the simplest approaches failed to explain
the spatial distribution of plankton populations but also more sophisticated
models were unable to account consistently for the most remarkable
features~\cite{powellokubo}. Even now, it is not clearly understood
why under apparent homogeneous conditions, such as temperature and
nutrients, plankton is still patchilly distributed. This particular
situation is of remarkable importance since in the absence of external
sources of patchiness the pattern must arise as a mere consequence
of the interactions between the individuals. The typical form of these
patterns is illustrated in Fig.~\ref{fig1}. The main trait is that
zooplankton is more patchilly distributed than phytoplankton~\cite{levi1,mackas1}.

The most intuitive model that can be proposed to explain plankton
dynamics~\cite{murray} considers the population densities of prey
(phytoplankton), \( N \), and predators (zooplankton), \( P \):
\begin{equation}
\label{prey}
{\partial N\over \partial t}=F_{N}(N,P)+D_{N}\nabla ^{2}N\; \; ,
\end{equation}
\begin{equation}
\label{predator}
{\partial P\over \partial t}=F_{P}(N,P)+D_{P}\nabla ^{2}P\; \; ,
\end{equation}
 where \( D_{N} \) and \( D_{P} \) are diffusion coefficients; and
\( F_{N} \) and \( F_{P} \) are functions that account for the interaction
between both species. This class of models is the most frequently
used in theories on pattern formation in ecology~\cite{murray,may3,may1}.
In the case of plankton, they were able to display spatial heterogeneity
under homogeneous conditions~\cite{levi3}. Zooplankton, however,
was less patchilly distributed than phytoplankton, in contradiction
with the observed pattern~\cite{levi1,levi3}.

There are two relevant features that are not taken into account by
this kind of models. \textit{First}, diffusion in the sea is not quantitatively
well modeled by usual Fickian diffusion~\cite{okubobook}. Both types
of diffusion processes will tend to spread and mix the populations,
but the specific form in which this is achieved is different. \textit{Second},
there is always an intrinsic stochasticity associated with the dynamics
of the population~\cite{mont,levindurrett}. From birth to death,
all processes share some degree of chance. The way randomness manifests
in the dynamics of the individuals depends on the scale we are looking
at~\cite{levi1}; deterministic equations are expected to be valid
in the limit of high numbers of individuals~\cite{vanK}. Therefore,
a deterministic description may be a reasonable one for phytoplankton
alone, but this does not need to be so for zooplankton which has much
fewer individuals~\cite{copewhite}. More importantly, while phytoplankton
interacts mainly with zooplankton, zooplankton interacts also with
fish and whales which are far from being evenly distributed. 

These two additional features have been incorporated in a prey-predator
model: \begin{equation}
\label{preytdn}
{\partial N\over \partial t}=F_{N}(N,P)-\vec{v}\cdot \vec{\nabla }N\; ,
\end{equation}
\begin{equation}
\label{predatortdn}
{\partial P\over \partial t}=F_{P}(N,P)-\vec{v}\cdot \vec{\nabla }P+\xi (t)\; ,
\end{equation}
 where dispersal is given by advection with a velocity field \( \vec{v} \)
{[}\( \equiv \vec{v}(\vec{r}) \){]} that depends on the position
\( \vec{r} \), and where a noise term \( \xi (t) \) has been included.

In general, the effects of the advective terms depend on the precise
form of the velocity field. For some turbulent fields~\cite{majdakramer},
the effect of advection can be simplified as follows: given a passive
field \( f(\vec{r},t) \) which evolves as \begin{equation}
{\partial f\over \partial t}=-\vec{v}\cdot \vec{\nabla }f\; .
\end{equation}
 the spatial Fourier transform of \( f(\vec{r},t) \) follows from
\begin{equation}
{df_{k}\over dt}=-D|k|^{\beta }f_{k}\; ,
\end{equation}
 where \( k \) is the wave number and \( D \) a constant. In this
case, advection can effectively be viewed as a diffusion process with
a diffusion coefficient \( D_{eff}(k)=D|k|^{2-\beta } \) that depends
on the scale. In contrast to usual Fickian diffusion, the variance
of the field is not proportional to \( t \) but is given by \( \left< r^{2}\right> \sim t^{2/\beta } \).
This is the type of time dependence observed for the dispersion of
tracers in the sea~\cite{okubobook,okubodia}, from which one can
obtain the explicit value of the parameter \( \beta  \).

To render our model analytically tractable, we consider the system
around a stable state. Fluctuations in zooplankton, \( \xi  \), move
the system away from equilibrium. If the fluctuations are not too
large, we can perform a linear expansion of \( F_{P} \) and \( F_{N} \):
\begin{eqnarray}
F_{N}(N,P) & = & c_{N}-a_{11}N-a_{12}P\; ,\nonumber \\
F_{P}(N,P) & = & c_{P}+a_{21}N-a_{22}P\; .\label{linear} 
\end{eqnarray}
 Here \( c_{N} \), \( c_{P} \), \( a_{11} \), \( a_{12} \), \( a_{21} \),
and \( a_{22} \) are positive constants. For the simplest form of
the noise term~\cite{vanK}, Gaussian white and uncorrelated in space
{[}\( \left< \xi (\vec{r},t)\right> =0 \) and \( \left< \xi (\vec{r},t)\xi ({\vec{r}}^{\prime },t^{\prime })\right> =2\sigma ^{2}\delta ({\vec{r}}^{\prime }-\vec{r})\delta (t^{\prime }-t) \){]},
and for \( a_{22}\sim 0 \) the variance spectra are given by \begin{eqnarray}
S_{N}(k)={a_{12}^{2}\sigma ^{2}\over (\tilde{D}_{N}+\tilde{D}_{P})\tilde{D}_{N}\tilde{D}_{P}}\; \; \; \mbox {and}\; \; S_{P}(k) & = & {\sigma ^{2}\over \tilde{D}_{P}}\; .
\end{eqnarray}
 where \( \tilde{D}_{N}\equiv D|k|^{\beta }+a_{11} \), \( \tilde{D}_{P}\equiv D|k|^{\beta }+a_{22} \),
and \( \sigma ^{2} \) is the intensity of the noise source. The assumptions
involved do not substantially constrain the applicability of the results.
When \( a_{22} \) is not negligible, the expressions become more
involved but the qualitative behavior is still the same. In particular,
the high wave-number limit remains unchanged. On the other hand, the
type of noise we have considered is quite general and can arise, among
others, from a random distribution of predators feeding on zooplankton
or even from the birth process itself~\cite{nature2001}. Other types
of noise with different properties, e.g. as those induced by turbulence~\cite{majdakramer,powellokubo},
are certainly present but we assume that they are not relevant for
the spectral properties of the pattern at the mesoscales.

The variance spectra obtained from previous equations display a power-law
region with exponent \( {-3\beta } \) for the phytoplankton and \( {-\beta } \)
for the zooplankton. The value of \( \beta \sim0 .87 \) inferred
from diffusion in the sea~\cite{okubodia} leads to exponents \( -2.6 \)
and \( -0.8 \), both in excellent agreement with field data~\cite{levi1,varspec,denman}.
It is worth emphasizing that the power law behavior appears only for
sufficiently high wavenumber (short scales); for low wavenumbers (long
scales), the variance spectra is flat, as observed in most field data~\cite{levi1,varspec,denman}.

In the same way, one can compute the coherence between two patterns
at different times~\cite{denman}, which provides information about
the global dynamics. For the phytoplankton this quantity is given
by \begin{equation}
\label{squco}
{N(k,\Delta t)\over N(k,0)}={\tilde{D}_{P}e^{\tilde{D}_{N}\Delta t}-\tilde{D}_{N}e^{\tilde{D}_{P}\Delta t}\over \tilde{D}_{P}-\tilde{D}_{N}}\; \; ,
\end{equation}
 where \( N(k,\Delta t)=\int _{0}^{\infty }||N(k,\omega )||^{2}e^{-i\omega \Delta t}d\omega  \).
This result indicates that short scales lose their correlation faster
than long ones and that eventually the whole pattern will be decorrelated,
as observed in satellite measurements~\cite{denman}.

In Fig. \ref{fig2} we plot the typical form of the variance spectra
and the squared coherence for different time lags. Both of them are
in excellent agreement with field data~\cite{levi1,varspec,denman,perfect3}.
Remarkably, the main properties of the pattern already appear in the
linear regime. Therefore, nonlinear interactions that drive the system
towards a stable state will lead to similar results. To study this
aspect in more detail, we have performed numerical simulations for
typical nonlinear interactions as explained in the caption of Fig.
\ref{fig3}. The resulting two-dimensional spatial distribution, transects,
and variance spectra (shown in Fig. \ref{fig3}) agree with both the
linear model and field data. Other types of nonlinear interactions
--- e.g. different functional responses --- as well as different types
of noise --- e.g. acting on zooplankton growth rate --- also produce
similar results (data not shown).

Field observations indicate that the power law region of the variance
spectra and the value of the exponent of this power law are robust
properties of the system; i.e. these properties are present under
a wide variety of conditions. In our model, there is always a power
law region whose exponent does not depend on biotic factors but is
completely determined by the specific form in which turbulent diffusion
acts on the system. This provides a straightforward explanation of
the predominance of the observed exponents for the phytoplankton falling
between \( -3 \) and \( -2 \). These are the values that arise for
2D (\( \beta =1 \)) and 3D (\( \beta =2/3 \)) isotropic turbulence,
respectively~\cite{majdakramer}. In the sea, the value of this exponent
will depend on the particular situation, but it is reasonable to assume
that it will be between those of 2D and 3D isotropic turbulence, as
the available data shows~\cite{okubobook,okubodia}. There are also
non-robust properties, such as the region where the variance spectra
turns flat. In the model, this depends on many factors: e.g. growth
and death rate, and turbulence. Field data shows that, indeed, the
position of this region exhibits great variability and that sometimes
it is not even present in the range of scales observed.

Turbulent diffusion and noise are two obvious features that have already
been considered in the context of marine ecosystems, but none of them
by itself has been able to explain the mesoscale patterns. In particular,
it is well known that noise generates variability, i.e. that noise
can be a source of patchiness~\cite{noisepatchiness,SR}. For instance,
reaction-diffusion prey-predator models with noise produce patterns
that at a glance strongly resemble those observed in the sea~\cite{steelemodel,ball}.
The exponents obtained (\( -6 \) for the phytoplankton and \( -2 \)
for the zooplankton), however, are far from the observed ones. This
quantitative, but not qualitative, disagreement is due to the dependence
of the effective diffusion coefficient with the scale. Thus, reaction-diffusion
models are unable to integrate correctly the scale dependence of the
physical properties of the environment. When this is taken into account,
noise not only generates patterns but is also able to produce the
right ones.

Turbulence plays a somehow ambivalent role. It can act in the same
way as diffusion does (transferring variance from smaller to larger
scales) and also in the opposite way (from larger to smaller scales).
These two types of processes are referred to as turbulent diffusion
and turbulent stirring, respectively. The former is the one we have
considered in our model. It was already considered in Ref.~\cite{powellokubo}
together with the type of noise that turbulence induces but without
the noise that can arise from biotic factors. The latter only plays
an important role when some degree of environmental heterogeneity
is present~\cite{tennekes}. Indeed, it has been shown that turbulent
stirring can generate patterns that resemble the observed ones if
spatial heterogeneity and time delays are introduced in the model~\cite{abrahanmodel}.
The type of time delays introduced, however, can lead non-realistic
situations such as growing zooplankton in the absence of phytoplankton.

Finally, it is important to emphasize that in our case noise is the
key element that allows moving from the individual to the population
description. Our results suggest that zooplankton dynamics at lower
scales affects the pattern at the mesoscale in the same way as noise
does. Considering a more detailed description is not necessary to
explain and to understand the main characteristics of the pattern.
This does not mean that the actual dynamics of zooplankton is not
important at all: its growth rate, its survival, and the intensity
of noise itself depend, among other factors, on how zooplankton aggregates
and on how it avoids its predators~\cite{levinind}. \emph{}It rather
means that under a wide range of conditions all those intricate mechanisms
will lead to patterns with properties as those induced by noise.

~\\
\textbf{Acknowledgments}\\
 This work was supported in part by the DGICYT of the Spanish Government.

{\raggedright \newpage\par}

\noindent \textbf{\large Figure Captions}{\large \par}

\begin{list}{FIGURE \arabic{figure}:}{\usecounter{figure}
\leftmargin 0 cm
\rightmargin 0 cm
\itemindent 1.5cm
}

\item \label{fig1} Transects obtained from field observations for
(a) phytoplankton and (b) zooplankton (redrawn from Ref.~\cite{mackas1}).
Here Chlorophyll \textit{a} is a measure of phytoplankton activity. 

\item \label{fig2} (a) Variance spectra \( S(k) \) for prey and
predators from Eqs. (\ref{prey}), (\ref{predator}), and (\ref{linear}).
(b) Square coherence \( [N(k,\Delta t)/N(k,0)]^{2} \) for prey {[}Eq.
(\ref{squco}){]} for time lags of \( 1 \), \( 6, \) and \( 7 \)
days. The values of the parameters are \( D=12 \), \( a_{11}=0.3 \),
\( a_{22}=0.05 \), and \( \beta =1 \). The length and time units
are Km and days, respectively. 

\item \label{fig3} Numerical simulations for a prey-predator model
given by Eqs. (\ref{prey}) and (\ref{predator}), with \( F_{N}(N,P)\equiv rN\left( 1-{N/K}\right) -cPf(N) \)
and \( F_{P}(N,P)=P\left( gf(N)-\epsilon \right)  \). The functional
response is \( f(N)=N^{2}/(1+N^{2}) \). Here, \( K \), \( c \),
\( g \), \( r \) and \( \epsilon  \) are positive constants. The
velocity field \( \vec{v}\equiv \vec{v}(\vec{r}) \) consists of a
series of vortices distributed as in Ref.~\cite{eddymodel} and the
noise is assumed to be Gaussian with zero mean and correlation function
\( \left< \xi (\vec{r},t)\xi ({\vec{r}}^{\prime },t^{\prime })\right> =2[\sigma P(\vec{r},t)]^{2}\delta ({\vec{r}}^{\prime }-\vec{r})\delta (t^{\prime }-t) \).
Typical transects for (a) prey and (b) predators. (c) Variance spectra
for prey and predators. These results were obtained by discretizing
the corresponding equations on a \( 250\times 250 \) two-dimensional
mesh~\cite{press}, with periodic boundary conditions and then by
using a standard method for integrating stochastic differential equations~\cite{kloeden}.
The values of the parameters are \( r=0.3 \), \( K=4 \), \( c=2 \),
\( g=0.1 \), \( \varepsilon =0.05 \), and \( \sigma =3.5 \). The
length and time units are Km and days, respectively. The size of each
cell of the discretized mesh is \( 0.25\times 0.25 \) Km\( ^{2} \).

\end{list}
\newpage\pagestyle{empty}~\vfill

\vspace{0.3cm}
{\centering \resizebox*{8cm}{!}{\includegraphics{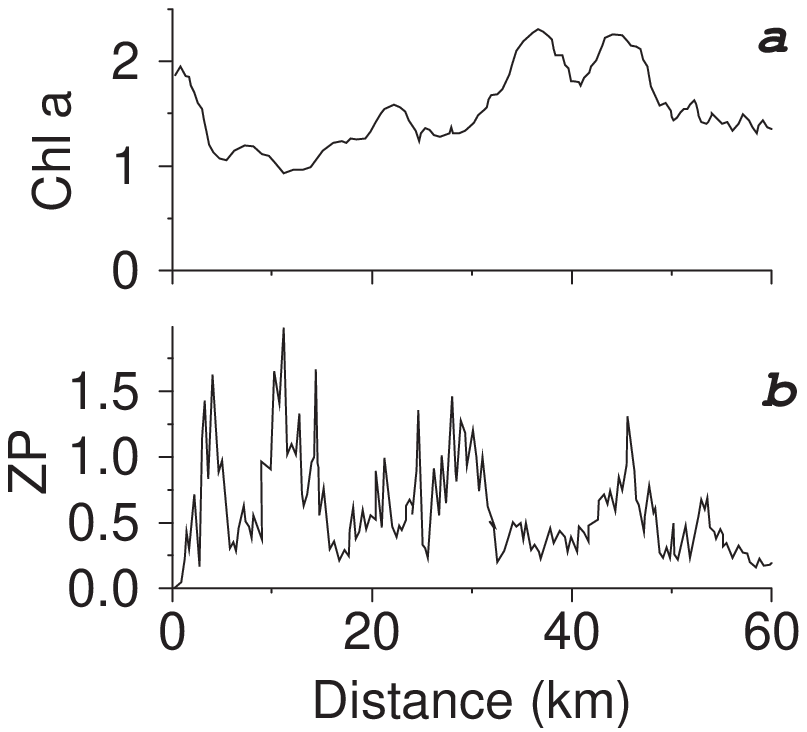}} \par}
\vspace{0.3cm}

 \vfill

~\textbf{FIGURE 1}

\newpage\pagestyle{empty}~\vfill

\vspace{0.3cm}
{\centering \resizebox*{7cm}{!}{\includegraphics{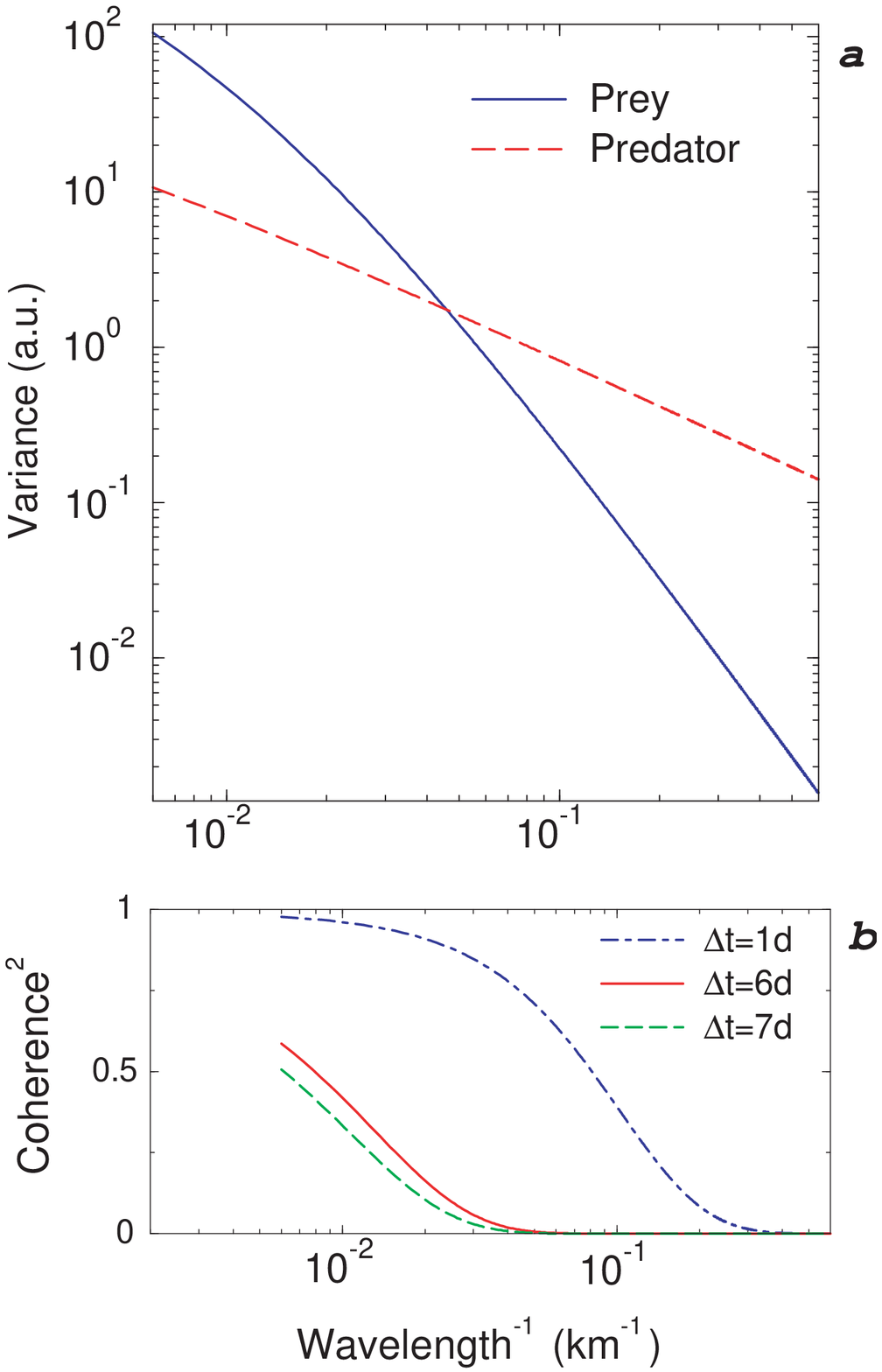}} \par}
\vspace{0.3cm}

 \vfill

~\textbf{FIGURE 2}

\newpage\pagestyle{empty}~\vfill

\vspace{0.3cm}
{\centering \resizebox*{14cm}{!}{\includegraphics{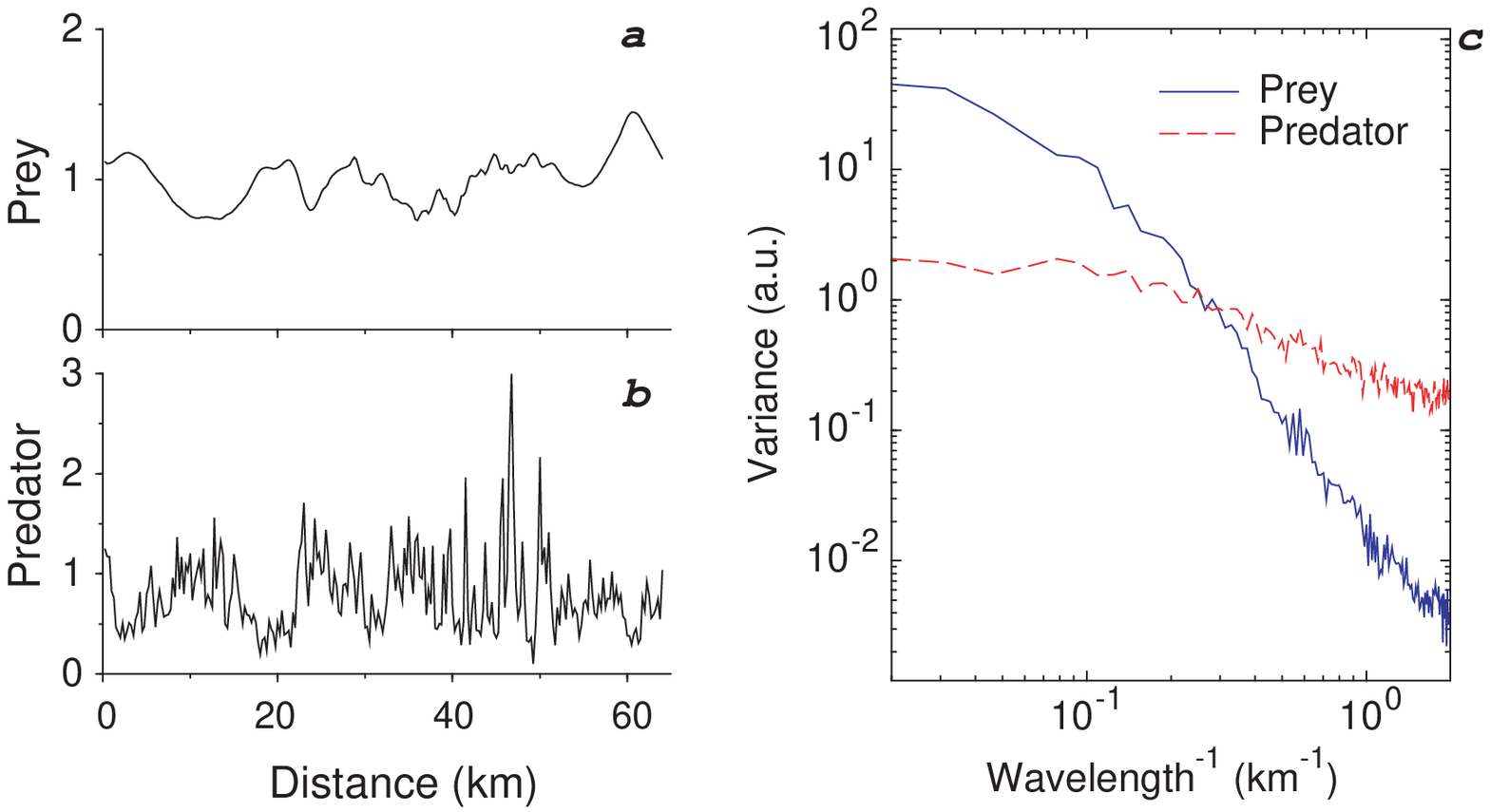}} \par}
\vspace{0.3cm}

 \vfill

~\textbf{FIGURE 3}
\end{document}